\documentclass[pra,twocolumn,superscriptaddress,preprintnumbers,eqsecnum,amsmath,amssymb]{revtex4}
\usepackage{graphicx}
\usepackage{dcolumn}
\usepackage{bm}
\usepackage{color}
\usepackage{epsfig}
\newcommand{\qcf}{quantum characteristic functions }
\newcommand{\Qcf}{Quantum characteristic functions }
\newcommand{\cf}{characteristic functions }
\newcommand{\beq}{\begin{eqnarray}}
\newcommand{\eeq}{\end{eqnarray}}
\newcommand{\al}{\alpha}

\newcommand{\lan}{\langle}
\newcommand{\ran}{\rangle}
\newcommand{\hz}{\hbar\rightarrow 0}

\newcommand{\Tr}{{\text Tr}}
\newcommand{\e}{{\text e}}
\newcommand{\rmd}{{\text d}}

\def\blambda{{\mbox{\boldmath $\lambda$}}}
\def\bmu{{\mbox{\boldmath $\mu$}}}
\input xy 
\xyoption{all}

\begin{document}

\title{Equations of Motion for the Quantum Characteristic Functions}
\author{Amir Kalev}
\email{amirk@techunix.technion.ac.il}
\affiliation{Department of Physics, Technion -- Israel Institute of Technology, Haifa 32000,
Israel.}
\author{Itay Hen}  \email{itayhe@post.tau.ac.il}
\affiliation{Raymond and Beverly Sackler School of Physics and Astronomy,
Tel-Aviv University, Tel-Aviv 69978,
Israel.
}%

\date{\today}  

\begin{abstract}
In this paper, we derive equations of motion for the normal-order, the symmetric-order and
the antinormal-order quantum characteristic functions,
applicable for general Hamiltonian systems.
We do this by utilizing the `characteristic form' of both quantum states and Hamiltonians.
The equations of motion we derive here are rather simple in form and in essence,
and as such have a number of attractive features.
As we shall see, our approach enables the descriptions of quantum and classical time evolutions
in one unified language. It allows for a direct comparison
between quantum and classical dynamics, providing
insight into the relations between quantum and classical behavior,
while also revealing a smooth transition between quantum
and classical time evolutions. In particular, the $\hz$ limit of the quantum equations of motion
instantly recovers their classical counterpart.
We also argue that the derived equations may prove to be very useful in numerical simulations.
\end{abstract}

\maketitle
\section{Introduction}
\Qcf  play an important role
in quantum optics, serving as alternative descriptions of quantum
states -- descriptions which are advantageous over other equivalent formalisms
in certain situations, for instance, in the evaluation
of expectation values of operator moments \cite{Lou}.
\Qcf are also very useful
in describing the dynamics of open systems such as the damped harmonic
oscillator, the quantum Brownian particle, two-level atoms and lasers (for reviews, see for
example \cite{Barnett,Carm,Qnoise}). For these kinds of systems, one
usually reformulates the appropriate master equation
(which involves Hilbert space operators) in terms of characteristic functions,
to obtain a differential Fokker-Planck equation,
whose methods of solution have been very well studied \cite{Lou}.
Recently, \qcf have been shown to
be appealing from an experimental point of view as well, providing
valuable tools for highly sensitive measurements \cite{Richter}.
\par
While the use of \cf for the description of quantum states
has a long history, to the best of our knowledge, a general equation of
motion (EOM) to describe their dynamics has not yet been formulated \cite{exception}.
This is with the exception of \cite{Gu},
where a general EOM for somewhat modified quantum characteristic functions
has been constructed, using a group-theoretical approach.

In this paper, we suggest a novel
derivation of the EOMs for the normal-order, symmetric-order and antinormal-order
\qcf which makes use of the characteristic form
of both quantum states and Hamiltonians.
As we shall see, this approach provides several attractive features:
Firstly, the derivation of the equations is both very general and
straightforward. It leads to EOMs which are rather simple in form; they involve
neither Hilbert-space operators nor infinite sums (as opposed to, e.g.,
the EOM for the Wigner function),
making them very appealing both for analytical calculations
as well as for practical purposes such as numerical simulations.

Moreover, as we shall see, in their final
form the EOMs turn out to be very similar to their
classical counterpart. This property avails insight
into the relation between quantum and classical time evolutions.
Furthermore, the formulation of quantum and classical evolutions
in a unified language enables the evaluation of the classical limit in a straightforward manner.
These issues are discussed in greater detail later on.
\par
The paper is organized as follows.
In Sec.~\ref{sec:fund} we recall the definition
of the quantum characteristic functions, and review some of their fundamental properties.
In Sec.~\ref{sec:eom}, we carry out the derivation of the EOMs,
and in Sec.~\ref{sec:clasLim} the relation of the equations
to their classical analogue is discussed.
We shall conclude with a few comments.

\section{\label{sec:fund}The quantum characteristic functions}
Three kinds of \qcf
have been widely used in the literature \cite{rem1}.
These are the normal-order, symmetric-order, and antinormal-order functions,
defined, respectively, by
the following mappings of density matrices $\hat{\rho}$ to complex-valued functions \cite{Lou}
\begin{subequations} \label{eq:PWQtilde}
\beq \label{eq:Ptilde}
C^{(\textrm{n})}(\lambda,\mu)&=& \Tr[\hat{\rho}\, \e^{i \bar{\xi} \hat{a}^{\dagger}} e^{i \xi \hat{a}}] \,, \label{P}\\
C^{(\textrm{s})}(\lambda,\mu)&=& \Tr[\hat{\rho}\, \e^{i \xi \hat{a}+i \bar{\xi} \hat{a}^{\dagger}}] \,, \label{W}\\
C^{(\textrm{a})}(\lambda,\mu)&=& \Tr[\hat{\rho}\, \e^{i \xi \hat{a}} \e^{i \bar{\xi} \hat{a}^{\dagger}}] \,.\label{Q}
\eeq
\end{subequations}
Here, $\xi$ is a complex variable defined
by the real coordinates $(\lambda,\mu)$ as
\beq
\xi= \sqrt{\frac{\hbar}{2 \omega}} \left(   \lambda - i \omega \mu\right) \,,
\eeq
and the  operators $\hat{a}$ and $\hat{a}^{\dagger}$ are annihilation and creation
operators related to the usual position  and  momentum operators by
\beq
\hat{x}&=& \sqrt{\frac{\hbar}{2 \omega}} (\hat{a}^{\dagger} +\hat{a}) \,,  \\\nonumber
\hat{p}&=& i \sqrt{\frac{\hbar \omega}{2}}  (\hat{a}^{\dagger} -\hat{a}) \,.
\eeq
The relations between the three characteristic functions are given by
\beq \label{eq:RelP}
C^{(\textrm{n})}(\lambda,\mu)=\e^{\frac1{2}|\xi|^2} C^{(\textrm{s})}(\lambda,\mu)=\e^{|\xi|^2}C^{(\textrm{a})}(\lambda,\mu)
\,,
\eeq
and they satisfy accordingly the following bounds
\begin{subequations}
\beq \label{eq:PQWbound}
\vert C^{(\textrm{n})}(\lambda,\mu) \vert &\leq& \e^{\frac1{2}|\xi|^2} \,, \\
\vert C^{(\textrm{s})}(\lambda,\mu) \vert &\leq& 1 \,, \\
\vert C^{(\textrm{a})}(\lambda,\mu) \vert &\leq& \e^{-\frac1{2}|\xi|^2} \,.
\eeq
\end{subequations}
It will be useful for our purpose to
also recall the inverse of (\ref{eq:Ptilde}), which maps the normal-order characteristic functions
to density matrices
\beq\label{eq:AqOp}
\hat{\rho}&=&\frac1{4 \pi^2} \int \rmd x \rmd p \int \rmd \lambda \rmd \mu \, \e^{-i(\lambda x + \mu p)} C^{(\textrm{n})}(\lambda,\mu)
 | \alpha \ran \lan \alpha | \,.\nonumber \\
\eeq
Here, $|\al \ran$ is a coherent state, where
the complex variable $\al$ is defined by the canonical coordinates $(x,p)$ via
\beq
\al = \frac1{\sqrt{2 \hbar \omega}} (\omega x +i p) \,.
\eeq
We also note that the
normal-order, the symmetric-order, and the antinormal-order characteristic functions
(\ref{eq:PWQtilde})
may also serve as the definitions of the Glauber-Sudarshan  $P$-representation \cite{PdistGla},
the Wigner  function \cite{Wigner}, and the Husimi $Q$ quasi-probability distribution \cite{Qdist}, respectively,
by means of a Fourier transform.
\section{\label{sec:eom}Equations of motion for the quantum characteristic functions}
The time evolution of an arbitrary quantum state $\hat{\rho}$
whose dynamics is governed by some Hamiltonian  $\hat{H}$ is given by
the von Neumann equation \beq \label{EOM1}
\frac{\partial}{\partial t} \hat{\rho} = \frac{i}{\hbar}
\left(\hat{\rho} \hat{H}-\hat{H} \hat{\rho} \right) \,. \eeq
In what follows, we rewrite this equation in terms of the normal-order
characteristic function $C^{(\textrm{n})}(\lambda,\mu)$. We shall first consider one-dimensional systems;
the generalization to multiple dimensions as well as
the formulation of the equation in terms of the  symmetric and
antinormal characteristic functions will be discussed later.
\par
Using Eq.~(\ref{eq:AqOp}), the left-hand-side of the von Neumann equation
is simply rewritten as
\beq
&& \frac{\partial}{\partial t} \hat{\rho} =  \\\nonumber
&& \frac1{4 \pi^2} \int \rmd x \rmd p \int \rmd \lambda \rmd \mu \, \e^{-i(\lambda x + \mu p)} \frac{\partial C^{(\textrm{n})}(\lambda,\mu)}{\partial t}
 | \alpha \ran \lan \alpha | \,.
\eeq
As for the right-hand-side, we make use of the fact that quantum observables, specifically Hamiltonians,
may also be formally represented in characteristic form \cite{Carm}.
In the normal-order case, this representation reads
\beq\label{eq:AqOp2}
\hat{H}&=&\frac1{2\pi \hbar} \int \rmd x \rmd p \int \rmd \lambda \rmd \mu \, \e^{-i(\lambda x + \mu p)} H^{(\textrm{n})}(\lambda,\mu)
 | \alpha \ran \lan \alpha | \,, \nonumber \\
\eeq
with the inverse transformation given by
\beq \label{eq:Adelta}
2 \pi H^{(\textrm{n})}(\lambda,\mu)= \hbar \, \Tr[\hat{H}\, \e^{i \bar{\xi} \hat{a}^{\dagger}} \e^{i \xi \hat{a}}] \,.
\eeq
In terms of  $C^{(\textrm{n})}(\lambda,\mu)$ and $H^{(\textrm{n})}(\lambda,\mu)$, the product $\hat{\rho} \hat{H}$
in the von Neumann equation  thus becomes
\begin{widetext}
\beq \label{multip}
\hat{\rho} \hat{H}= \frac1{8 \pi^3 \hbar}
\int \rmd \lambda_1 \rmd \mu_1 \rmd \lambda_2 \rmd \mu_2  \int \rmd x_1 \rmd p_1 \rmd x_2 \rmd p_2 C^{(\textrm{n})}(\lambda_1,\mu_1)
H^{(\textrm{n})}(\lambda_2,\mu_2)
\e^{-i(\lambda_1 x_1 + \mu_1 p_1 +\lambda_2 x_2 + \mu_2 p_2 )} \lan \alpha_1 | \alpha_2 \ran
 | \alpha_1 \ran \lan \alpha_2 |  \,.\nonumber\\
\eeq
\end{widetext}
Next, we identify the characteristic representation
of $\hat{\rho} \hat{H}$, by expressing the operator
$| \alpha_1 \ran \lan \alpha_2 |$  in Eq. (\ref{multip}), in a coherent state diagonal form
according to the identity \cite{Scully}
\begin{equation}
| \alpha_1 \ran \lan \alpha_2 |=\int \rmd^2 \alpha \, \Tr \big[| \alpha_1 \ran \lan \alpha_2| \hat{\Delta}(\alpha,\bar{\alpha}) \big] \,
| \al \ran \lan \al | \,,
\end{equation}
where
\beq
\hat{\Delta}(\alpha,\bar{\alpha}) &=& \frac1{\pi^2}
\int \rmd^2 \alpha' \e^{-\alpha'(\bar{\alpha}-\hat{a}^{\dagger})}
\e^{ \bar{\alpha}' (\alpha-\hat{a})}\,.
\eeq
Integrating over the internal variables, Eq. (\ref{multip}) simplifies to
\beq
\hat{\rho} \hat{H} &=& \int \rmd x \rmd p \int \rmd \lambda \rmd \mu
\, C^{(\textrm{n})}_{\rho H}(\lambda,\mu) \,
\e^{-i(\lambda x +\mu p)}  |\al \rangle \langle \al| \,,\nonumber\\
\eeq
with $C^{(\textrm{n})}_{\rho H}(\lambda,\mu)$ -- the normal-order characteristic function of the  product
$\hat{\rho} \hat{H}$ -- given by \beq C^{(\textrm{n})}_{\rho
H}(\lambda,\mu) &=&
\int \rmd \lambda' \rmd \mu' C^{(\textrm{n})}(\lambda',\mu') H^{(\textrm{n})}(\lambda-\lambda',\mu-\mu') \nonumber \\
&\times& \e^{\frac{\hbar}{2 \omega }(\lambda'+i \omega \mu')(\lambda-\lambda'-i\omega (\mu-\mu'))} \,.
\eeq
Similarly, we find that the
normal-order characteristic function of the product $\hat{H} \hat{\rho}$ is given
by \beq C^{(\textrm{n})}_{H \rho}(\lambda,\mu)
&=&
\int \rmd \lambda' \rmd \mu' C^{(\textrm{n})}(\lambda',\mu') H^{(\textrm{n})}(\lambda-\lambda',\mu-\mu') \nonumber \\
&\times& \e^{\frac{\hbar}{2 \omega }(\lambda'-i \omega \mu')(\lambda-\lambda'+i\omega (\mu-\mu'))}
\,. \eeq
Substituting
the expressions for $\hat{\rho} \hat{H}$ and $\hat{H} \hat{\rho}$
into the von Neumann equation and equating the integrands on each side,
we arrive at the final form
\beq\label{QEOM1d}
\frac{\partial}{\partial t} C^{(\textrm{n})}(\lambda,\mu) &=& \int \rmd
\lambda' \rmd \mu' K^{(\textrm{n})}(\lambda,\mu, \lambda',\mu')
 \\\nonumber &\times&
H^{(\textrm{n})}(\lambda-\lambda',\mu-\mu')
\, C^{(\textrm{n})}(\lambda',\mu') \,,
\eeq
with the `normal-order kernel' $K^{(\textrm{n})}(\lambda,\mu, \lambda',\mu')$ identified as
\beq\label{eq:Kn}
&&K^{(\textrm{n})}(\lambda,\mu, \lambda',\mu') \\\nonumber
&&=\frac{2}{\hbar}
\e^{
\frac{\hbar}{2 \omega} \left[
\lambda'(\lambda -\lambda') + \omega^2 \mu' (\mu - \mu')
\right]
} \sin
\frac{\hbar}{2}(\lambda \mu' -\mu \lambda')
\,.
\eeq
Eq.(\ref{QEOM1d}) is the EOM for one-dimensional normal-order
characteristic functions.
\par
Of course, in cases where the dynamics of the system under consideration is
governed by a Hamiltonian consisting of a linear combination of products of annihilation
and creation operators, Eq. (\ref{QEOM1d}) can be further simplified.
This is due to the fact the characteristic representation of Hamiltonians of the above form
are linear combinations of delta functions and derivatives thereof \cite{Carm},
so the right-hand-side of (\ref{QEOM1d}) may be immediately integrated to produce
a linear partial differential equation for $C^{(\textrm{n})}(\lambda,\mu)$,
to be further solved by the usual methods.
\par
The EOMs for the symmetric-order $C^{(\textrm{s})}(\lambda,\mu)$ and the antinormal $C^{(\textrm{a})}(\lambda,\mu)$
are readily obtained by making use of the relations (\ref{eq:RelP})
both for $C^{(\textrm{n})}(\lambda,\mu)$ and for $H^{(\textrm{n})}(\lambda,\mu)$.
The resultant equations have the same form as Eq. (\ref{QEOM1d}), but with
different kernels. The antinormal-order kernel is
\beq \label{eq:Ka}
&&K^{(\textrm{a})}(\lambda,\mu, \lambda',\mu') \\\nonumber
&& =
\frac{2}{\hbar}
\e^{-
\frac{\hbar}{2 \omega} \left[
\lambda'(\lambda -\lambda') + \omega^2 \mu' (\mu - \mu')
\right]
} \sin
\frac{\hbar}{2}(\lambda \mu' -\mu \lambda')
\,,
\eeq
whereas the symmetric-order kernel turns out to be particularly simple and reads
\beq \label{eq:Ks}
K^{(\textrm{s})}(\lambda,\mu, \lambda',\mu') =
\frac{2}{\hbar} \sin
\frac{\hbar}{2}(\lambda \mu' -\mu \lambda')
\,.
\eeq
The simple form of the symmetric kernel enables further simplification
of the `symmetric-order' EOM (Eq. (\ref{QEOM1d}) with $\scriptstyle{(n)} \to \scriptstyle{(s)}$).
It allows one to Fourier transform the EOM and to obtain an EOM
for the Wigner function $W(x,p)$ -- the Fourier transform of $C^{(\textrm{s})}(\lambda,\mu)$.
Expanding the kernel $K^{(\textrm{s})}$ in a Taylor series in $\hbar$,
the right-hand-side of the symmetric EOM becomes
\beq \label{eq:QEOMw1}
& \sum_{m=0}^{\infty} & b_m \int \rmd \lambda' \rmd \mu'
\bigg[ \mu' (\lambda-\lambda')-\lambda' (\mu-\mu') \bigg]^{2 m+1}
 \nonumber\\
&\times&  H^{(\textrm{s})}(\lambda-\lambda',\mu-\mu') \, C^{(\textrm{s})}(\lambda',\mu') \,,
\eeq
with $\displaystyle{b_m=(-1)^m \frac{(\hbar/2)^{2m}}{(2m+1)!}}$.
Further expanding the bracketed term above in a binomial series,
and Fourier transforming the resultant convolutions
term by term, the EOM becomes
\begin{widetext}
\beq \label{eq:QEOMw3}
 \frac{\partial }{\partial t}W(x,p)=
 \sum_{m=0}^{\infty} \sum_{k=0}^{2 m +1}  b_m (-1)^k{2m+1 \choose k}i^{2(2m+1)}
\left( \frac{\partial^k}{\partial x^k}\frac{\partial^{2m+1-k}}{\partial p^{2m+1-k}}W(x,p)\right)
 \, \left( \frac{\partial^{2m+1-k}}{\partial x^{2m+1-k}}\frac{\partial^{k}}{\partial p^{k}}
 \tilde{H}^{(\textrm{s})}(x,p)\right)  \,,
\eeq
\end{widetext}
where $\tilde{H}^{(\textrm{s})}(x,p)$ is the Fourier transform of
$H^{(\textrm{s})}(\lambda,\mu)$. Eq. (\ref{eq:QEOMw3})
is thus a generalized EOM for the Wigner function under general Hamiltonian dynamics.
\par
This EOM may be written in a compact form with the help of the following
notation:
\beq
\frac{\partial^{\scriptscriptstyle{(1)}}}{\partial p}
\frac{\partial^{\scriptscriptstyle{(2)}}}{\partial x}AB \equiv
\left(\frac{\partial}{\partial p}A\right)\,\left(\frac{\partial}{\partial x}B\right)\,,
\eeq
where $A$ and $B$ are arbitrary functions of the canonical variables.
Via this definition, Eq. (\ref{eq:QEOMw3}) simplifies to
\beq
\label{eq:QEOMw}
&&\frac{\partial}{\partial t}W(x,p)\\\nonumber
&=&\sum_{m=0}^{\infty}  b_m  \left(\frac{\partial^{\scriptscriptstyle{(1)}}}{\partial p}
 \frac{\partial^{\scriptscriptstyle{(2)}}}{\partial x}-\frac{\partial^{\scriptscriptstyle{(1)}}}{\partial x}
 \frac{\partial^{\scriptscriptstyle{(2)}}}{\partial p}\right)^{2m+1}W(x,p)\tilde{H}^{(\textrm{s})}(x,p)\\\nonumber
&=&\frac{2}{\hbar} \sin
\frac{\hbar}{2}\left(\frac{\partial^{\scriptscriptstyle{(1)}}}{\partial p}
\frac{\partial^{\scriptscriptstyle{(2)}}}{\partial x}-
\frac{\partial^{\scriptscriptstyle{(1)}}}{\partial x}
\frac{\partial^{\scriptscriptstyle{(2)}}}{\partial p}\right)W(x,p)\tilde{H}^{(\textrm{s})}(x,p)\,.
\eeq
In the special case where $\tilde{H}^{(s)}(x,p)=p^2/2 + V(x)$,
Eq. (\ref{eq:QEOMw}) reduces to the well-known equation for the time evolution of the
Wigner function \cite{Schleich}
\beq
\frac{\partial}{\partial t}W(x,p)&=& \Bigg(
\frac{\rmd V(x)}{\rmd x} \frac{\partial}{\partial p} -p \frac{\partial}{\partial x}
\\\nonumber
&+& \sum_{m=0}^{\infty} \frac{\rmd^{2 m+1} V(x)}{\rmd x^{2 m+1}} \frac{\partial^{2 m+1}}{\partial p^{2 m+1}} \Bigg) W(x,p)
\,.
\eeq
\par
The generalization of the EOMs derived above to $N$ dimensions is straightforward.
Now, points in phase space are represented by vector pairs
\hbox{$(\mathbf{x},\mathbf{p})=(x_1,\cdots,x_N,p_1,\cdots,p_N)$},
and their Fourier counterparts by \hbox{$(\blambda,\bmu)=(\lambda_1,\cdots,\lambda_N,\mu_1,\cdots,\mu_N)$}.
In an exact analogy with the one-dimensional case, it is an easy matter to check that
the EOM for the $N$-dimensional characteristic function $C^{(\textrm{n})}(\blambda,\bmu)$ is
\beq \label{eq:QEOMcNdim}
\frac{\partial C^{(\textrm{n})}(\blambda,\bmu)}{\partial t} &=&
\int \rmd \blambda' \rmd \bmu'
 K^{(\textrm{n})}(\blambda,\bmu,\blambda',\bmu')
 \\\nonumber
&\times&  H^{(\textrm{n})}(\blambda-\blambda',\bmu-\bmu') \, C^{(\textrm{n})}(\blambda',\bmu') \,,
\eeq
with
\beq \label{eq:KqNdim}
&&K^{(\textrm{n})}(\blambda,\bmu, \blambda',\bmu') \\\nonumber
&& =
\frac{2}{\hbar}
\e^{
\frac{\hbar}{2 \omega} \left[
\blambda'(\blambda -\blambda') + \omega^2 \bmu' (\bmu - \bmu')
\right]
} \sin
\frac{\hbar}{2}(\blambda \bmu' -\bmu \blambda')
\,,
\eeq
while the EOMs for the symmetric-order $C^{(\textrm{s})}(\blambda,\bmu)$ and the antinormal $C^{(\textrm{a})}(\blambda,\bmu)$
are obtained by making use of the multi-dimensional form
of relations (\ref{eq:RelP}):
\beq \label{eq:Rel2}
C^{(\textrm{n})}(\blambda,\bmu)&=&\e^{\frac{\hbar}{4 \omega}(
\blambda^2+\omega^2 \bmu^2)}C^{(\textrm{s})}(\blambda,\bmu)\nonumber \\
&=&
\e^{\frac{\hbar}{2 \omega}(
\blambda^2+\omega^2 \bmu^2)}C^{(\textrm{a})}(\blambda,\bmu)
\,,
\eeq
with similar relations for $H^{(\textrm{n})}(\blambda,\bmu)$.
The antinormal-order kernel then becomes
\beq \label{eq:KaNdim}
&&K^{(\textrm{a})}(\blambda,\bmu, \blambda',\bmu') \\\nonumber
&& =
\frac{2}{\hbar}
\e^{-
\frac{\hbar}{2 \omega} \left[
\blambda'(\blambda -\blambda') + \omega^2 \bmu' (\bmu - \bmu')
\right]
} \sin
\frac{\hbar}{2}(\blambda \bmu' -\bmu \blambda')
\,,
\eeq
whereas the symmetric-order kernel is
\beq \label{eq:KsNdim}
K^{(\textrm{s})}(\blambda,\bmu, \blambda',\bmu') =
\frac{2}{\hbar} \sin
\frac{\hbar}{2}(\blambda \bmu' -\bmu \blambda')
\,.
\eeq

\section{\label{sec:clasLim}Relation to classical dynamics}
Next, we show that the von Neumann equation, in its form (\ref{eq:QEOMcNdim}),
enables a direct comparison between quantum and classical time evolutions.
To see this, let us write down the corresponding classical EOM,
namely the Liouville equation \cite{vanKampen}, in terms of
classical characteristic functions.
As with the quantum equation, we start off with a general
one-dimensional system
and generalize to multiple dimensions later.
\par
The time evolution of a statistical distribution $P(x,p)$ evolving in time under a
Hamiltonian $H(x,p)$
is given by the Liouville equation
\beq \label{eq:clasEOMP}
\frac{\partial P}{\partial t}
=\frac{\partial P}{\partial x} \frac{\partial H}{\partial p} -
\frac{\partial P}{\partial p} \frac{\partial H}{\partial x} \,.
\eeq
In terms of the characteristic functions of $P(x,p)$ and $H(x,p)$, defined by the
Fourier transforms
\begin{subequations}
\label{PHclas}
\beq
P(x,p) &=& \int \rmd \lambda \rmd \mu
C^{(\textrm{c})}(\lambda,\mu) \e^{-i (\lambda x + \mu p)} \,, \label{Pclas} \\
H(x,p) &=& \int \rmd \lambda \rmd \mu
H^{(\textrm{c})}(\lambda,\mu) \e^{-i (\lambda x + \mu p)} \,,
\eeq
\end{subequations}
the Liouville equation becomes
\beq
\frac{\partial C^{(\textrm{c})}(\lambda,\mu)}{\partial t} &=&
\int \rmd \lambda' \rmd \mu'
K^{(\textrm{c})}(\lambda,\mu,\lambda',\mu') \\\nonumber
&\times&  H^{(\textrm{c})}(\lambda-\lambda',\mu-\mu') \, C^{(\textrm{c})}(\lambda',\mu')\,,
\eeq
where the `classical kernel' $K^{(\textrm{c})}(\lambda,\mu,\lambda',\mu')$ is given by
\beq
K^{(\textrm{c})}(\lambda,\mu,\lambda',\mu')=\lambda \mu' - \mu \lambda' \,.
\eeq
This can be readily verified by substituting (\ref{PHclas}) into (\ref{eq:clasEOMP}).
In $N$ dimensions the equation generalizes to
\beq \label{eq:EOMcNdim}
\frac{\partial C^{(\textrm{c})}(\blambda,\bmu)}{\partial t} &=&
\int \rmd \blambda' \rmd \bmu'
K^{(\textrm{c})}(\blambda,\bmu,\blambda',\bmu') \\ \nonumber
&\times& H^{(\textrm{c})}(\blambda-\blambda',\bmu-\bmu') \, C^{(\textrm{c})}(\blambda',\bmu') \,,
\eeq
with
\beq \label{eq:KcNdim}
K^{(\textrm{c})}(\blambda,\bmu,\blambda',\bmu') = \blambda \cdot \bmu' - \bmu \cdot \blambda' \,.
\eeq
\par
Having the quantum EOM (\ref{eq:QEOMcNdim})
and the classical EOM (\ref{eq:EOMcNdim}) on an equal footing,
the difference between quantum and classical time evolutions
becomes transparent. This difference is encapsulated in the dissimilarity
between the quantum and classical kernels;
this is with the understanding that in both cases the Hamiltonian
characteristic function is the same.
\par
Moreover, comparing the symmetric-order kernel with the classical kernel,
we obtain the simple relation
\beq \label{eq:KqwKc}
K^{(\textrm{s})}=
\frac{2}{\hbar} \sin
\frac{\hbar}{2} K^{(\textrm{c})}
\,.
\eeq
Relation (\ref{eq:KqwKc}) reveals a smooth transition between quantum time evolution
and classical evolution, one in which $\hbar$ naturally plays the role
of a `quantumness' parameter. Specifically, in the $\hbar \to 0$ limit of $K^{(\textrm{s})}$,
we obtain the fully classical behavior, namely
$\lim_{\hbar \to 0} K^{(\textrm{s})} =K^{(\textrm{c})}$.
As one would expect, the $\hbar \to 0$ limit of normal-order and the antinormal-order
kernels also recover the classical kernel $K^{(\textrm{c})}$.

\section{Summary and further remarks}
In formulating the von Neumann equation
as an integro-differential equation for each of the three quantum characteristic functions,
we have obtained a `classical' description for the time evolution of quantum systems,
which further enabled the treatment of quantum
and classical systems on an equal footing.
As a bonus we have also obtained a generalized EOM for the Wigner function.
In the new formulation, the difference between the quantum and classical EOMs narrows
down to the dissimilarity between the respective kernels.
Moreover, the correspondence between quantum and classical systems arises
in this formulation very naturally; a straightforward evaluation of the $\hz$ limit of the quantum
kernels recovers the classical one.
This new formulation may thus become very useful
when one is interested in contrasting the behavior of quantum systems
with their parallels in the classical world, as analogies can be drawn between
classical phenomena and phenomena generated by quantum dynamics.
In fact, a similar strategy has recently proved to be very helpful
in the context of connecting the no-broadcasting theorem with its classical counterpart \cite{KH}.
\par
We believe that the  EOM presented here, may prove to be useful for practical
reasons as well. The reason for that is twofold. Firstly,
the quantum characteristic functions are scalar-valued and are ensured to be well-behaved
(in the case of the symmetric and the
antinormal functions -- also bound); when numerical simulations are concerned,
these properties are of utmost importance, as they allow computations
to take place in a rather straightforward manner and without having to worry
about divergences and singularities in the evolving systems. This is in contrast with
the Glauber-Sudarshan  $P$-representation \cite{PdistGla} for example, which is ill-defined
for some perfectly reasonable quantum states and is thus very problematic in this respect.
Secondly, the equation derived here is `classical' in nature; it involves
neither Hilbert-space operators nor infinite sums.
This property makes the equation very convenient for simulation purposes,
as this type of equations is fairly simple to implement computationally.
\begin{acknowledgments}
We thank Venketeswara Pai and Ady Mann for useful suggestions and discussions.
\end{acknowledgments}


\begin{thebibliography}{99}
\bibitem{exception}
Equations of motion in terms of characteristic functions have been derived
for several particular open systems, such as the ones noted earlier.
\end{thebibliography}
\end{document}